\documentclass[prd,superscriptaddress,floatfix,nofootinbib,twocolumn]{revtex4-2}
\pdfoutput=1 
\usepackage[T1]{fontenc} 
\usepackage{amsmath,amssymb,amsfonts,amsthm}
\usepackage{graphicx}
\usepackage[usenames,dvipsnames]{color}
\usepackage{enumitem}
\usepackage{verbatim}
\usepackage[percent]{overpic}
\usepackage{rotating}
\usepackage[colorlinks=true,allcolors=black]{hyperref}
\usepackage{array}
\usepackage{mathtools}
\usepackage{lmodern}
\usepackage[normalem]{ulem}
\usepackage{braket}
\usepackage{tensor}
\usepackage{dsfont}
\usepackage{bm}
\usepackage{tikz}

\usepackage{mathrsfs}

\usetikzlibrary{decorations.pathmorphing}
\usepackage{CJKutf8}

\definecolor{patriarch}{rgb}{0.5, 0.0, 0.5}

\newcommand{\x}{\mathsf{x}}

\usetikzlibrary{positioning}
\usetikzlibrary{calc,through,backgrounds}

\theoremstyle{definition}
\newtheorem{definition}{Definition}[section]
\newtheorem{theorem}{Theorem}[section]

\newcommand{\kako}[1]{\left( #1 \right)}
\newcommand{\kagikako}[1]{\left[ #1 \right]}
\newcommand{\ts}[1]{ _{\text{#1}} }

\allowdisplaybreaks

\newcommand{\dd}{\text{d}}

\newcommand{\sx}{\mathsf{x}}

\newcommand{\ii}{\mathsf{i}}

\begin{document} 

\title{More Excitement Across the Horizon}

\author{Mar\'ia R. Preciado-Rivas}
\email{mrpreciadorivas@uwaterloo.ca} 
\affiliation{Institute for Quantum Computing, University of Waterloo, Waterloo, Ontario, N2L 3G1, Canada}
\affiliation{Department of Applied Mathematics, University of Waterloo, Waterloo, Ontario, N2L 3G1, Canada}

\author{Manar Naeem}
\email{manar.naeem@uwaterloo.ca}
\affiliation{Institute for Quantum Computing, University of Waterloo, Waterloo, Ontario, N2L 3G1, Canada}
\affiliation{Department of Physics and Astronomy, University of Waterloo, Waterloo, Ontario, N2L 3G1, Canada}

\author{Robert B. Mann}
\email{rbmann@uwaterloo.ca}
\affiliation{Institute for Quantum Computing, University of Waterloo, Waterloo, Ontario, N2L 3G1, Canada}
\affiliation{Department of Physics and Astronomy, University of Waterloo, Waterloo, Ontario, N2L 3G1, Canada}
\affiliation{Perimeter Institute for Theoretical Physics,  Waterloo, Ontario, N2L 2Y5, Canada}

\author{Jorma Louko}
\email{jorma.louko@nottingham.ac.uk}
\affiliation{School of Mathematical Sciences, University of Nottingham, Nottingham NG7 2RD, UK} 

\date{February 2024; updated June 2024.\\ aaPublished in Phys.\ Rev.\ D \textbf{110}, 025002 (2024), doi.org/10.1103/PhysRevD.110.025002.\\ aaFor Open Access purposes, this Author Accepted Manuscript is made available under CC BY public copyright.}

\begin{abstract}
An Unruh-DeWitt detector falling radially into a four-dimensional Schwarzschild black hole, coupled linearly to a massless scalar field that has been prepared in the Hartle-Hawking or Unruh state, has recently been shown to exhibit a local extremum in its transition probability near the black hole horizon [K.K.~Ng et al., New J.\ Phys.\ {\bf 24} (2022) 103018]. We show that a similar phenomenon is present in the transition rate of an Unruh-DeWitt detector falling radially into a spinless Ba\~nados-Teitelboim-Zanelli (BTZ) black hole, with the scalar field prepared in the Hartle-Hawking state. We give extensive numerical results as a function of the detector's energy gap, the black hole's mass, and the detector's drop-off radius. Our results suggest that the effect is robust, motivating a search for a similar effect in other black hole spacetimes, and calling for an explanation of the physical origin of the effect.
\end{abstract}

\maketitle
\flushbottom

\section{Introduction}

While a quantum theory of gravity continues to elude us, the field of relativistic quantum information has emerged as a powerful tool for investigating quantum effects in curved spacetime. 
Within this framework, models for particle detectors provide us with an operational approach for probing quantum fields in scenarios with no distinguished notion of particle. 
The simplest model is the Unruh-DeWitt (UDW) particle detector~\cite{Unruh.effect, DeWitt1979},  which is a two-level quantum system that locally interacts with a quantum scalar field. 
One quantity of interest is the transition probability between its two levels; another is the derivative of this probability with respect to the total detection
time, known as the transition rate. One well-known result obtained from this model is that a uniformly linearly accelerated detector experiences the Minkowski vacuum as thermal, with a temperature proportional to the proper acceleration, a phenomenon known as the Unruh effect \cite{Unruh.effect,DeWitt1979,Davies1975}.  Another result is the degradation of the entanglement between two detectors that are in relative non-inertial motion \cite{Fuentes-Schuller:2004iaz,Alsing:2006cj}.

In a similar manner, one can examine the response function of a detector in black hole spacetimes.
For example, a static detector in the exterior of a  Schwarzschild black hole responds thermally to the Hartle-Hawking(-Israel) state~\cite{HartleHawking1976,Israel1976}; similarly,
the transition rate of a stationary detector co-rotating with the Ba\~nados-Teitelboim-Zanelli (BTZ) black hole thermalizes at the Hawking temperature~\cite{hodgkinsonStaticStationaryInertial2012}.

Much less is known about the response of detectors that freely fall into a black hole and cross its horizon. 
The general expectation appears to be that the detector's transition rate will be smooth and the detector will not thermalize. 
An adiabatic formalism for estimating the effective temperature has been developed in \cite{Barbado:2011dx,Barcelo:2010pj,Barbado:2012pt,Smerlak:2013sga,Barbado:2016nfy}.
A detector freely falling through a stationary cavity in a Schwarzschild background was shown to probe spacetime curvature when compared to that of an equivalently accelerated detector travelling through an inertial cavity in the absence of curvature \cite{Ahmadzadegan:2013iua}. Other studies have taken place in more simplified lower-dimensional settings.
A numerical study of the transition rate of a UDW detector freely falling toward a $(1+1)$ dimensional Schwarzschild black hole provided evidence of how its thermal properties are gradually lost during the infall \cite{Juarez-Aubry:2014jba,JuarezPhDThesis}, 
and the transition rate of a detector infalling toward a Cauchy horizon in $(1+1)$ dimensions was shown to diverge for a field in both the Unruh and Hartle-Hawking-Israel states~\cite{Juarez-Aubry:2021tae}; in both cases the detector was taken to couple to the momentum of the field, to cure infrared ambiguities that occur in $(1+1)$ dimensions. Coupling to field momentum was also employed in a study of the behaviour of entanglement and mutual information
acquired by two freely falling detectors in   $(1+1)$ dimensional Schwarzschild spacetime, finding that 
correlations can be harvested even when the detectors are causally disconnected by the event horizon \cite{Gallock-Yoshimura:2021yok}. 
In $(2+1)$ dimensions, the transition rate of a detector 
radially falling toward a static 
BTZ black hole was computed \cite{hodgkinsonStaticStationaryInertial2012}, but the detector was switched on and off in the region exterior to the black hole; no regions in parameter space were found where the transition rate was thermal.

Recently, the response of  a detector freely falling toward and across the horizon of a $(3+1)$-dimensional Schwarzschild black hole was considered 
\cite{Ng2022}. 
Taking the detector's initial position to be at infinity with zero initial velocity, the response function, a multiple of the transition probability, was 
numerically calculated for a field in the Hartle-Hawking(-Israel) and Unruh states.  Surprisingly, a small but discernible local extremum in the response appears as the detector crosses the horizon of the black hole for both field states. The locus of the peak or dip in the response function depends on the interaction duration and the detector's energy gap.

It is of significant interest to understand the extent to which this phenomenon takes place for other black holes. To this end, we investigate the transition rate (the temporal derivative of the response function) of a freely falling UDW detector into a static BTZ black hole in $(2+1)$ dimensions. This setting affords several advantages. First, it extends previous work~\cite{hodgkinsonStaticStationaryInertial2012} on freely falling detectors in this setting, allowing us to calibrate the detector's transition rate at large distances from the horizon with that across the horizon.  More importantly, the response can be computed as a sum over images instead of the mode sum necessary in the Schwarzschild case \cite{Ng2022}, a considerable technical advantage that allows us to explore a broader range of parameter space.

We find that the BTZ case exhibits similar, but much richer, behaviour to that of the Schwarzschild case \cite{Ng2022}.  The transition rate exhibits the general feature of slow oscillation as a function of proper time 
as the detector  approaches the black hole. However, near and beyond the horizon the rate sharply peaks and then drops  before rapidly increasing near the black hole's singularity. Furthermore, there are specific times at which the transition rate is not differentiable -- we refer to these as ``glitches''.  Both features depend on the mass of the black hole and so serve as a kind of `early warning system' as to whether the detector will continue
to freely fall in AdS$_3$ spacetime or  across an event horizon in a spacetime of the same local geometry. 

The outline of our paper is as follows. In section~\ref{sec:setup} we present the basic formalism needed to compute the transition rate of a detector freely falling in both the BTZ spacetime and an AdS$_3$ spacetime of the same local geometry. We present in section~\ref{sec:results} the results of our study, illustrating with a number of figures  how the transition rate depends on the mass of the black hole, the initial conditions of the trajectory,  the energy gap of the detector, and  the boundary conditions of the field at asymptotic infinity.  In section~\ref{sec:conclusions}, we sum up our work, drawing conclusions about the results and suggesting new avenues for future study.

\section{Setup}
\label{sec:setup}

We consider the Unruh-DeWitt model for a detector coupled to a scalar field.  Since much of the formalism has been previously presented~\cite{hodgkinsonStaticStationaryInertial2012}, we shall only recapitulate the essential tools needed for our study.

A UDW detector is a pointlike qubit having  two energy levels, denoted by 
$|g\rangle$ and~$|e\rangle$, 
with the respective energy eigenvalues $0$ and~$E$.
The quantity $E$
may be positive
or negative; 
 in the former case case $|g\rangle$ is the ground state and $|e\rangle$ is the excited state, while in the latter case the roles are reversed.
  The detector  moves on a timelike worldline~$\x(\tau)$, 
parametrized by  its proper time $\tau$ and interacts with a real massless scalar field~$\phi$ 
via the interaction Hamiltonian 
\begin{equation}
H_{\text{int}}= \lambda \chi(\tau)\mu(\tau)\phi\bigl(\x(\tau)\bigr) 
\ , 
\end{equation}
where  $\mu(\tau)
=
        \ket{e} \bra{g} e^{ \ii E \tau }
        +
        \ket{g} \bra{e} e^{ -\ii E \tau }$ is 
the detector's monopole moment operator, 
$\lambda$ is a  coupling constant, and
$\chi$~is the switching function, specifying 
how the detector is switched on and off.

 In first-order perturbation theory, the probability of the detector to make a transition from the state $|g\rangle$ to the state $|e\rangle$ is proportional to the response function, given by 
\begin{equation}\label{Fnodot}
    \mathcal{F} = \int \mathrm{d}\tau' \, \mathrm{d}\tau'' \, 
    \mathrm{e}^{-\ii E(\tau' - \tau'')}W(\tau',\tau'') \,,
\end{equation}
where $W$ is the pullback of the field's Wightman function in the state in which the field was initially prepared \cite{Kay:1988mu,Fewster:1999gj,Junker:2001gx,Louko:2007mu}. 
In $(2+1)$ spacetime dimensions, we may take $\chi$ to be the characteristic function of a time interval, since $\mathcal{F}$ remains well defined for this $\chi$ despite the nonsmoothness at the switch-on and switch-off moments, 
and we may further consider the detector's transition rate, given by (a multiple of) the derivative of $\mathcal{F}$ with respect to the switch-off moment. The transition rate can be written as \cite{hodgkinsonStaticStationaryInertial2012}
\begin{equation}\label{Fdot}
    \dot{\mathcal{F}}_\tau = \frac{1}{4} + 2 \int^{\Delta\tau}_{0}\,\mathrm{d}s \operatorname{Re}\left[ \mathrm{e}^{-\ii Es}W(\tau,\tau-s)\right] \,,
\end{equation}
where $\tau$ denotes the switch-off moment and $\Delta\tau$ denotes the total proper time that the detector operates.
The transition rate has a measurable interpretation in terms of an ensemble of identical detectors, all following the trajectory $\sx(\tau)$ and switched off at distinct moments 
\cite{Louko:2007mu,Satz:2006kb}.

We are interested in the transition rate of a detector on a radially in-falling geodesic in a spinless nonextremal BTZ black hole spacetime~\cite{BTZ1, BTZ2}. 
Outside the horizon, the line element is 
\begin{align}
\label{btzmet}
    \dd s^2
    &=
        - f(r) \dd t^2
        + \dfrac{ \dd r^2 }{f(r)}
        + r^2 
        \dd \varphi^2\,,
\end{align} 
where $f(r)=r^2/\ell^2  -M$, $t\in \mathbb{R}$,  $\varphi \in [0, 2\pi)$, the dimensionless parameter $M$ is the mass of the black hole, where  
 nonextremality means $M>0$. 
The exterior is at 
$r\in (r_h,\infty)$, where $r=r_h \equiv \ell \sqrt{M}$
is the location of the event horizon.
The metric \eqref{btzmet} is a vacuum solution to the Einstein field equations with cosmological constant $\Lambda=-1/\ell^2$, where $\ell\; ( > 0)$ is the anti-de Sitter (AdS) length. 

By  setting $\varphi \to y \in (-\infty,\infty)$ 
one obtains AdS$_3$-Rindler spacetime, where 
the parameter $M$ in \eqref{btzmet} can be se to unity without loss of generality \cite{Jennings:2010vk,Henderson:2019uqo}. 
Conversely,
the BTZ spacetime can be obtained from AdS$_3$-Rindler by identifying $ y \to \phi \in (0 ,2\pi\sqrt{M})$, and then rescaling $\phi\to \sqrt{M}\varphi$ and the radial and time coordinates to obtain \eqref{btzmet}; its local geometry is therefore 
 that of anti-de Sitter 
spacetime.
This fact will be important when one considers a quantum field in BTZ spacetime, in which the correlation functions can be written as a sum of correlators in AdS$_3$. 
 
\begin{figure}
    \centering
    \includegraphics[width=0.5\linewidth]{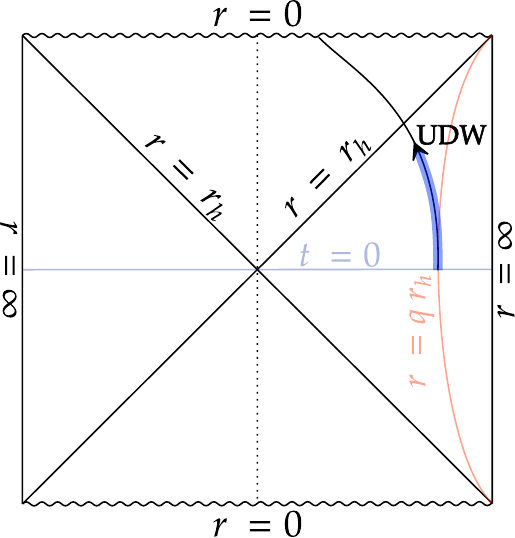}
    \caption{Penrose diagram indicating the trajectory followed by a UDW detector free falling into a BTZ black hole}
    \label{fig:diagram}
\end{figure}

The radial timelike geodesic followed by the detector is given in BTZ coordinates by \cite{hodgkinsonStaticStationaryInertial2012,carlipDimensionalBlackHole1995}
\begin{equation}
    \begin{aligned}
        t & =\left( \ell/\sqrt{M}\right) \operatorname{arctanh}{\left( \frac{\tan{\tilde{\tau}}}{\sqrt{q^2-1}}\right)},\\
        r & =\ell\sqrt{M}q\cos{\tilde{\tau}}, \\
        {\varphi} & =  {\varphi_0},\\
    \end{aligned}
    \label{eq:trajectory}
\end{equation}
where 
$q>1$, $\varphi_0$ is a constant value of $\varphi$, and $\tilde{\tau}$ is  a dimensionless affine parameter such that the proper time equals $\tilde{\tau}\ell$. $r$ reaches its maximum value $q\ell \sqrt{M}$ at $\tilde{\tau}=0$ with $t=0$, as depicted in Fig.~\ref{fig:diagram}. 

We  consider a massless conformally coupled quantum scalar field $\hat \phi(\sx)$ satisfying the Klein-Gordon equation, 
\begin{align}
    (\square - R/8) \hat \phi(\sx)=0\,,
\end{align}
where $\square$ is the d'Alembert operator and $R$ is the Ricci scalar. 
The Wightman function, in Eqs.\ \eqref{Fnodot} and \eqref{Fdot}, is the two-point correlation function 
in the vacuum state $\ket{0}$ of the field, $W(\sx, \sx')\coloneqq \braket{0|\hat \phi(\sx) \hat \phi(\sx')|0}$.
Since the BTZ spacetime is a $\mathbb{Z}$ quotient of an open set in AdS$_3$ \cite{BTZ1, BTZ2}, 
the BTZ field correlations can be calculated in terms of the 
Wightman function in AdS$_3$ spacetime by using the method of images \cite{LifschytzBTZ,carlipDimensionalBlackHole1995}:
\begin{align}
    &W\ts{BTZ}(\sx, \sx')
    =
        \sum_{n=-\infty}^\infty
        W\ts{ADS}(\sx, \Gamma^n \sx')
        \label{eq:BTZ Wightman} \\
    &=
        \dfrac{1}{ 4\pi \sqrt{2}\ell }
        \sum_{n=-\infty}^\infty
        \kagikako{
            \dfrac{1}{ \sqrt{ \sigma_\epsilon (\sx, \Gamma^n \sx') } }
            -
            \dfrac{\zeta}{ \sqrt{ \sigma_\epsilon (\sx, \Gamma^n \sx')+2 } }
        },
        \notag
\end{align}
where $\Gamma:(t,r,\varphi)\mapsto (t,r,\varphi+ 2\pi)$ in the exterior. 
The parameter $\zeta \in \{ -1,0,1\}$ specifies the boundary conditions of the field at asymptotic infinity: Neumann, transparent, or Dirichlet, respectively.
Dirichlet and Neumann boundary conditions at spatial infinity are imposed to take into account that
asymptotically AdS spacetimes (such as BTZ)
are  not globally hyperbolic. The boundary condition prevents information from escaping or leaking in, thereby allowing a good quantization scheme~\cite{LifschytzBTZ}.  Alternatively, the quantization can be carried out with transparent boundary conditions~\cite{Avis:1978}, but its physical meaning is not clear. For Dirichlet and Neumann boundary conditions, the BTZ Wightman function corresponds to a Kubo-Martin-Schwinger state outside the horizon and satisfies certain analyticity properties with respect to a vacuum state in Kruskal coordinates, characterizing a Hartle-Hawking state~\cite{LifschytzBTZ}.

The quantity
\begin{align}
    \sigma_\epsilon (\sx, \Gamma^n \sx')
    &= 
        \dfrac{r r' }{ r_h^2} 
        \cosh 
        \kagikako{
            \dfrac{r_h}{\ell} (\Delta \varphi - 2\pi n )
        }
        -1 \label{sigma} 
        \\
        &-\dfrac{ \sqrt{ (r^2-r_h^2) (r^{\prime 2}-r_h^2) } }{ r_h^2 } 
        \cosh 
        \kako{
            \dfrac{r_h}{\ell^2} 
            \Delta t - \ii \epsilon
\nonumber        }
\end{align}
is the squared geodesic separation (scaled by the square of the AdS length) between the two points in the covering space, 
with $\Delta \varphi\coloneqq \varphi - \varphi', \Delta t\coloneqq t-t'$, 
 and the notation encodes the distributional character of the Wightman function as the limit $\epsilon\to 0_+$.
Inserting \eqref{eq:BTZ Wightman} 
and \eqref{eq:trajectory}
into \eqref{Fdot}, we obtain
\cite{hodgkinsonStaticStationaryInertial2012} 
\begin{widetext}
\begin{multline}
        \dot{\mathcal{F}}_\tau(E)=\frac{1}{4}
        +\frac{1}{2 \pi \sqrt{2}} \sum_{n=-\infty}^{\infty} \int_0^{\Delta \tilde{\tau}} \mathrm{d} \tilde{s} \operatorname{Re}
        \left[\frac{\mathrm{e}^{-\ii \tilde{E} \tilde{s}}}{\sqrt{-1+K_n \cos \tilde{\tau} \cos (\tilde{\tau}-\tilde{s})+\sin \tilde{\tau} \sin (\tilde{\tau}-\tilde{s})}}\right. \\
        \left.-\zeta \frac{\mathrm{e}^{-\ii \tilde{E} \tilde{s}}}{\sqrt{1+K_n \cos \tilde{\tau} \cos (\tilde{\tau}-\tilde{s})+\sin \tilde{\tau} \sin (\tilde{\tau}-\tilde{s})}}\right],
        \label{eq:transition_rate}
\end{multline}
\end{widetext}
 where
\begin{equation}
    K_n := 1+2q^2\sinh^2{\left(n\pi\sqrt{M}\right)},
    \label{eq:K_n}
\end{equation}
with $\tilde{\tau}:= \tau/\ell$, $\Delta \tilde{\tau}:=\Delta \tau/\ell$, and $\tilde{E}:=E\ell$.
The detector is sharply switched on
at a time $\tau_0 = \tau - \Delta\tau$, where  $\tau$ corresponds to the moment it is sharply switched off
and observed. 

Several observations are in order. 

First, the $n\ne0$ terms in \eqref{eq:transition_rate} are invariant under $n\to-n$. The $n\ne0$ part of the sum hence reduces to twice the sum over $n\ge1$.

Second, while we have above arrived at the transition rate formula \eqref{eq:transition_rate} using the expression \eqref{sigma}, which is valid only in the black hole exterior, \eqref{eq:transition_rate} holds over the detector's full trajectory, even when the detector operates already before exiting the white hole and/or after entering the black hole. This follows from the existence of a global analytic chart, by analytic continuation in~\eqref{eq:transition_rate}.

Third, the transition rate decomposes as
\begin{equation}
\dot{\mathcal{F}}_\tau=\dot{\mathcal{F}}_\tau^{n=0}+\dot{\mathcal{F}}_\tau^{n \neq 0}
\end{equation}
where $\dot{\mathcal{F}}_\tau^{n=0}$ consists of the $n=0$ term,  and $\dot{\mathcal{F}}_\tau^{n \neq 0}$ consists of the sum over $n\ne0$ terms. $\dot{\mathcal{F}}_\tau^{n=0}$ depends only on $E\ell$ and $\Delta \tau / \ell$. That is, it does not depend  on the mass $M$ or the switch-on time $\tau_0$ (or switch-off time $\tau$) and is given by \cite{hodgkinsonStaticStationaryInertial2012}
\begin{equation}
    \dot{\mathcal{F}}_\tau^{n=0}(E)=\frac{1}{4}-\frac{1}{4\pi}\int^{\Delta\tilde{\tau}}_0\, d\tilde{s}
    \left[ \frac{\sin{\left(\tilde{E}\tilde{s}\right)}}{\sin{\left(\tilde{s}/2\right)}} + \zeta \frac{\cos{\left(\tilde{E}\tilde{s} \right)}}{\cos{\left( \tilde{s}/2\right)}}\right].
    \label{eq:transition_rate_n=0}
\end{equation}
which is the transition rate in AdS$_3$ spacetime.

Fourth, the square roots in \eqref{eq:transition_rate} are positive for positive arguments, and they are analytically continued to negative arguments by giving $\tilde{s}$ a small negative imaginary part~\cite{hodgkinsonStaticStationaryInertial2012,Kay:1988mu}. 
This follows from the distributional definition of the Wightman function, encoded in the $i\epsilon$ in~\eqref{sigma}. 

From now on we specialise to the case $\tau_0=0$, in which the detector is switched on at the moment when $r$ takes its largest value, as illustrated in Figure~\ref{fig:diagram}; 
the numerical results in the rest of the paper will be for this case. 
Then \eqref{eq:transition_rate} holds with 
$0<\tilde{\tau}<\pi/2$ and $\Delta\tilde{\tau}=\tilde{\tau}$, 
so that the integration range of $\tilde{s}$ is from $0$ to $\tilde{\tau}$. 
What remains is to find the branches of the square roots in the $n\ne0$ terms. In the part involving~$\zeta$, the function under the square root is positive for all $\tilde{s}$ in the integration interval, so the square root is positive. 
In the part not involving $\zeta$, the function under the square root is positive at $\tilde{s}=0$. It has in the integration interval at most one zero, denoted by~$\tilde{s}^*$, and the derivative of the function at $\tilde{s}^*$ is negative. It follows that the square root is positive for $\tilde{s} < \tilde{s}^*$ and positive imaginary for $\tilde{s}>\tilde{s}^*$. At $\tilde{s}=\tilde{s}^*$, the square root vanishes, so that the integrand is singular, but this singularity is of the inverse square root type and hence integrable, rendering the integral well defined. 

Fifth, the singularities in the integrand in \eqref{eq:transition_rate} have a geometric origin: they occur when the trajectory crosses the future light cone of the switch-on event. For pure AdS$_3$ ($n=0$), these crossings only occur for light cones reflected from infinity, and so they do not occur for the proper time durations we consider, as seen explicitly in \eqref{eq:transition_rate_n=0}. 
In the BTZ spacetime, however, such  crossings can occur, at arbitrarily many discrete values of the proper time, because the light cones wrap around the $\varphi$ direction. Any metric geometry in which the detector crosses the light cone of the switch-on event will have similar singularities in the transition rate integrand.

Finally, we note that when $\tau$ takes the discrete values for which
\begin{equation}
    \Delta\tilde{\tau}_n : = \arccos(1/K_n) \,, \ n=1,2,\ldots,
    \label{eq:glitches}
\end{equation} 
the integrand in \eqref{eq:transition_rate} has a term whose singularity coincides with an endpoint of the integration interval. This holds both for $\tau_0=0$ and more generally. Geometrically, \eqref{eq:glitches} means evaluating the transition rate at precisely those special moments where the trajectory crosses the lightcone of the switch-on event. The transition rate is still well defined at these special moments, but it is not differentiable in $\tau$; hence, at these special moments, the response function $\mathcal{F}$ in \eqref{Fnodot} is differentiable but not twice differentiable in~$\tau$. 
These ``kinks'' or ``glitches'' will feature prominently in our numerical results.

\section{Results}\label{sec:results}

In this section, we systematically arrange our results to facilitate the examination of the transition rate's dependence on various factors, namely the boundary conditions of the Wightman function at infinity given by $\zeta$, the detector energy gap  $E$,  the mass $M$ of the black hole, and the detector's initial position with respect to the black hole's event horizon, given by $q$. 
In all our computations, we
truncated the sum in \eqref{eq:transition_rate}   at the first term whose absolute value at $\Delta\tau/\ell = 1.57$ is smaller than $10^{-5}$. Taking into account the form of the individual contributions,   discussed below with reference to Fig.~\ref{fig:individual_contributions}, this truncation criterion suffices to ensure that the neglected terms are on the order of $10^{-5}$, or smaller, for all $\Delta\tau/\ell \in [0, 1.57]$.

\subsection{Transition rate for different boundary conditions}\label{sec:diff_bcs}

\begin{figure}[ht]
    \centering
    \includegraphics[width=\linewidth]{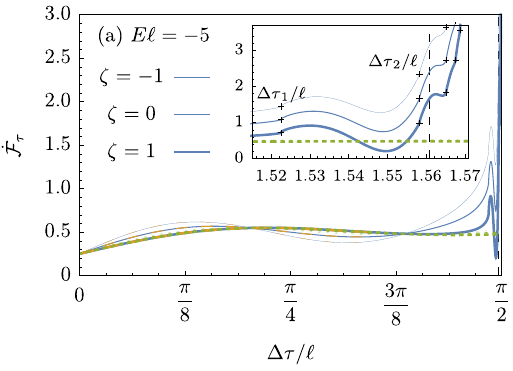}
    \includegraphics[width=\linewidth]{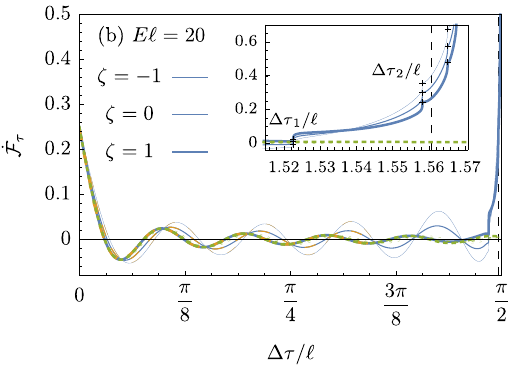}
    \caption{The transition rate $\dot{\mathcal{F}}_\tau$ of a detector freely falling into a BTZ black hole, shown by the blue, solid curves with $q=100$, $M=10^{-4}$, $\tau_0=0$, and (a) $E\ell=-5$ or (b) $E\ell=20$ calculated with $n=356$ or $n=320$ terms of the image sum, respectively. For comparison, we plot the AdS$_3$ case, given by the zeroth term $\dot{\mathcal{F}}_\tau^{n=0}$ (green, dotted curve).     
    We consider the Wightman function using Neumann (thinnest), transparent, and Dirichlet (thickest) boundary conditions. The results of \cite{hodgkinsonStaticStationaryInertial2012} are shown in orange dashed curves.
    The total time for the detector to reach the event horizon of the BTZ black hole is indicated with a dashed vertical line at $\Delta\tau/\ell=\arccos{(0.01)}$.
    The values of $\Delta\tau_n/\ell$ for $n=1,2,3$ are shown with crosses inside the insets, which are close-ups of the transition rate near the horizon crossing.
    }
    \label{fig:comparingtojorma}
\end{figure}

In Fig.~\ref{fig:comparingtojorma}, we show the transition rate $\dot{\mathcal{F}}_\tau$ as a function of the detector's total proper time $\Delta \tau /\ell$ for the three boundary conditions, $\zeta=-1$, 0 and 1, with $\tau_0=0$, $q=100$, $M=10^{-4}$, and energy gaps (a) $E\ell=-5$ or (b) $E\ell=20$. These parameters are the same as those used in the previous study of a freely falling detector in BTZ spacetime~\cite{hodgkinsonStaticStationaryInertial2012}
but (unlike that study) include the transition rate for $\Delta \tau /\ell\gtrsim\arccos{(1/q)}$, that is, near and beyond the horizon crossing. 
 For $\Delta \tau /\ell\lesssim\arccos{(1/q)}$, we find close agreement with the results of~\cite{hodgkinsonStaticStationaryInertial2012}, given by the orange dashed curves. 

Furthermore, we observe that the transition rate for a detection time shorter than the first glitch $\Delta\tau_1/\ell$ has a behaviour remarkably distinct from that after the first glitch. We first discuss the transition rate for $\Delta\tau/\ell\lesssim\Delta\tau_1/\ell$ (with $\Delta\tau_1/\ell\approx 1.52257$ for $q=100$ and $M=10^{-4}$).

\begin{figure}[t]
    \centering
    \includegraphics[width=\linewidth]{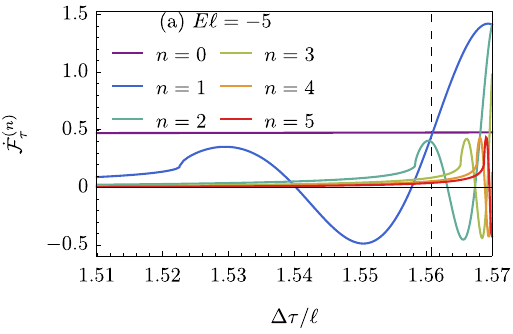}
    \includegraphics[width=\linewidth]{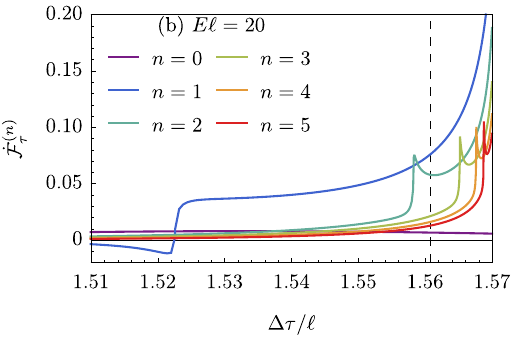}
    \caption{Individual contributions from the first six terms of the image sum for the transition rate calculated using the same parameters as Fig.~\ref{fig:comparingtojorma} and $\zeta=1$. The horizon crossing is shown with a dashed vertical line at  $\Delta\tau/\ell= \arccos(0.01) \approx  1.5608$.}
    \label{fig:individual_contributions}
\end{figure}

In the case of the Dirichlet boundary condition, $\zeta=1$, the transition rate bears a strong resemblance to the zeroth term 
 $\dot{\mathcal{F}}_\tau^{n=0}$, which
is the transition rate of a stationary detector in pure AdS$_3$.  This agreement is also present for other energy gaps  (as we illustrate in 
Figs.~\ref{fig:boundary_conditions_negative_gap} and \ref{fig:boundary_conditions_positive_gap} of the appendix
and is commensurate with previous work~\cite{hodgkinsonStaticStationaryInertial2012}).
 For the boundary conditions $\zeta=0$ and $-1$, by contrast, we observe a notable difference between the transition rate and the zeroth term $\dot{\mathcal{F}}_\tau^{n=0}$. 
This dissimilarity increases as the magnitude of the energy gap $|E\ell|$ decreases, becoming appreciable when $|E\ell|=5$ as seen in Figs.~\ref{fig:comparingtojorma}(a) and \ref{fig:boundary_conditions_positive_gap}(b).

We observe in Fig.~\ref{fig:comparingtojorma} 
(as well as in Figs.~\ref{fig:boundary_conditions_negative_gap} and \ref{fig:boundary_conditions_positive_gap} of the appendix) 
that the transition rate undergoes more oscillations as $|E\ell|$ increases. This characteristic is expected since $E\ell$ is the argument of a complex exponential function in Eq.~\eqref{eq:transition_rate}. We also note that the transition rates for $\zeta=-1$ and $\zeta=0$ are in-phase, whereas they are out-of-phase for $\zeta=1$. Another way we can describe the behaviour in this regime is that the $\zeta=0$ and $\zeta=-1$ boundary conditions have the effect of increasing the magnitude of the oscillations in the transition rate when compared to the oscillations in the rate calculated with $\zeta=1$.

We now turn our attention to the transition rate near and beyond the first glitch, which is shown in more detail in the insets of Fig.~\ref{fig:comparingtojorma}. The transition rate differs considerably from the 
AdS$_3$ case (the
zeroth term $\dot{\mathcal{F}}_\tau^{n=0}$) for detection times $\Delta\tau/\ell \gtrsim  \Delta\tau_1/\ell$, with the difference being more pronounced as more kinks $\Delta\tau_n/\ell$ are reached or the black hole singularity at $\pi/2$ is approached. To investigate this difference, we plot in Fig.~\ref{fig:individual_contributions} the individual contributions from the first few terms in the image sum using the same parameters as in Fig.~\ref{fig:comparingtojorma} and $\zeta=1$. We observe that beyond the $n$-th glitch, $\Delta \tau_n/\ell$, the contribution from the $n$-term rapidly becomes appreciable and comparable to the zeroth term, which gives rise to the divergence near the black hole singularity that we see in Fig.~\ref{fig:comparingtojorma}.  Note also that the $n=0$ term is very similar across the three boundary conditions; hence the difference in the transition rate stems from the contributions of the $|n|\ge1$ terms.

We also see in Fig.~\ref{fig:individual_contributions} that there is dissimilar behaviour in the contributions for negative and positive gaps. The terms for negative gaps are oscillatory with a pseudo-period depending on the magnitude of the gap, whereas the terms for positive gaps exhibit a dip right after the glitch but do not oscillate.
It is remarkable that the contribution of the $n$-th term oscillates when $\Delta\tau/\ell\lesssim\Delta\tau_n/\ell$ for any gap (not shown). However, the amplitude of these oscillations becomes negligible far from the glitch when compared to the amplitude of the zeroth term. This is as expected, considering that the transition rate in this interval resembles the pure AdS$_3$ term.

For $\Delta\tau/\ell \gtrsim \Delta\tau_1/\ell$, the contributions add up in such a way that the transition rate diverges 
 as $\Delta\tau/\ell \to \pi/2$.  
For $\zeta=0$ and $\zeta=-1$, the terms are large enough   that the transition rate for these boundary conditions has a positive deviation from the rate obtained with $\zeta=1$. This deviation becomes more pronounced as the total detection time approaches $\pi/2$,  as depicted in Fig.~\ref{fig:comparingtojorma} (and Figs.~\ref{fig:boundary_conditions_negative_gap} and \ref{fig:boundary_conditions_positive_gap} of the appendix). Overall, the transition rates are notably numerically different for the three types of boundary conditions, as previously 
observed~\cite{hodgkinsonStaticStationaryInertial2012}, but
the qualitative behaviour of the transition rate remains consistent across the various boundary conditions. Therefore, we confine the results presented in the main text hereafter to the case $\zeta=1$.

Of particular note is the transition rate in Fig.~\ref{fig:comparingtojorma}(a), which attains a prominent local maximum followed by a local minimum before the horizon crossing at $\Delta\tau/\ell=\arccos(0.01)$; that is, a peak and a dip appear at switch-off times for which the detector has not yet reached the event horizon. Specifically, the local maximum/minimum feature is in between the glitches $\Delta\tau_1/\ell$ and $\Delta\tau_2/\ell$.
In the close-up near the horizon crossing shown in the inset, we observe that the transition rate attains another, less prominent, local maximum followed by another local minimum. 
The aforementioned feature in the transition rate for a freely falling detector in a BTZ black hole resembles the local extremum that was observed in the Schwarzschild case \cite{Ng2022}.

\subsection{Transition rate for different energy gaps}\label{sec:diff_gaps}

In the previous subsection, we briefly discussed the effect of the detector's energy gap on the transition rate, considering the various boundary conditions. In this subsection, we delve into a more detailed discussion by exploring additional values of the energy gap.

\begin{figure*}[t]
    \centering
    \includegraphics[width=0.49\linewidth]{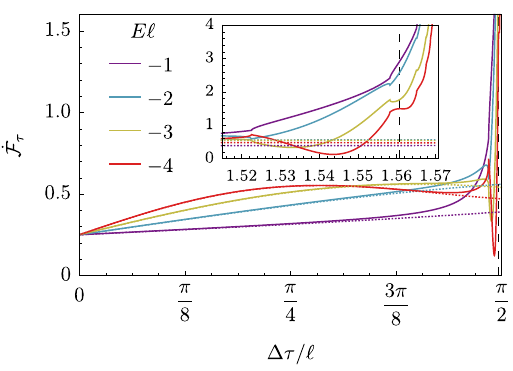}
    \includegraphics[width=0.49\linewidth]{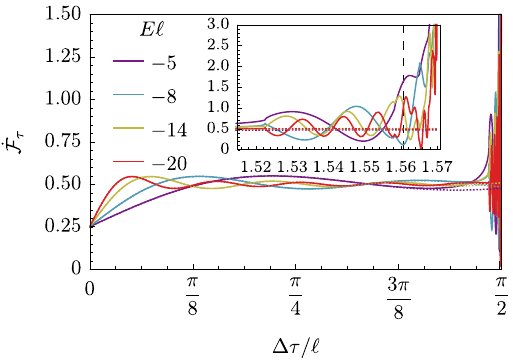}
    \includegraphics[width=0.49\linewidth]{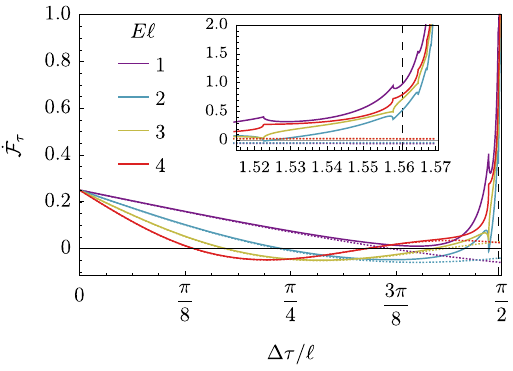}
    \includegraphics[width=0.49\linewidth]{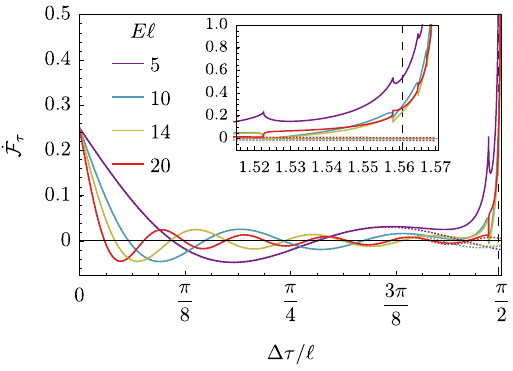}
    \caption{Transition rate $\dot{\mathcal{F}}_\tau$ calculated with $M=10^{-4}$, $q=100$, $\tau_0=0$, $\zeta=1$ and varying the detector's energy gap. The image sum for $E\ell=\pm1,\, \pm2,\, \pm3, \pm4,\, \pm5,\, -8,\, 10,\, \pm14$ and $\pm20$ is calculated with $n=388,\, 361,\, 333,\, 349,\, 356,\, 337,\, 333,\, 327$ and $320$ terms, respectively. The AdS case (zeroth term $\dot{\mathcal{F}}_\tau^{n=0}$) for each gap is shown with dotted curves of the same colour.}
    \label{fig:varying_gap_z_1}
\end{figure*}

In Fig.~\ref{fig:varying_gap_z_1}, we present the transition rate varying the detector's energy gap for fixed black hole mass $M=10^{-4}$, $q=100$, $\tau_0=0$ and $\zeta=1$. 
We observe that the transition rate starts increasing (decreasing) for $E\ell<0$ ($E\ell>0$), attaining a first local maximum (minimum) that is located at shorter $\Delta\tau/\ell$ for larger $|E\ell|$. As also noted in the previous subsection, the number of oscillations in $\Delta\tau/\ell\lesssim\Delta\tau_1/\ell\approx 1.52257$ that follow the first extremum increases when the absolute value of $E\ell$ increases, but the amplitude of these oscillations gradually diminishes. 
For these detection times, the transition rate oscillates around $\sim0.5$ for $E\ell<0$ and $\sim0$ for $E\ell>0$.

The results in Fig.~\ref{fig:varying_gap_z_1} are in accord with the intuition that a wider energy gap in an excited detector leads to a stronger tendency for decay to the ground state. Conversely, a narrower energy gap in a detector in the ground state makes it more susceptible to excitation.
The fact that the deexcitation transition rate remains positive across the entire detection time domain indicates that the response function is monotonically increasing with respect to the detection time. Conversely, the occurrence of a negative excitation transition rate within certain intervals of the detection time suggests that the response function exhibits oscillatory behaviour, displaying local extrema at the points where the transition rate crosses zero.

With respect to $\Delta\tau/\ell\gtrsim\Delta\tau_1/\ell$, we note that the behaviour of the transition rate for negative gaps differs significantly from that for positive gaps, as observed in the insets of Fig.~\ref{fig:varying_gap_z_1} (and Fig.~\ref{fig:individual_contributions} of the previous subsection). 
As the magnitude of $E\ell$ increases, the transition rate for $E\ell<0$ experiences a larger number of oscillations of diminishing amplitude for fixed $\Delta\tau/\ell$. On the other hand, for $E\ell>0$, 
 the transition rate
features dips or peaks in between the glitches $\Delta\tau_n/\ell$ but no appreciable oscillations.
We can trace this difference back to the sign of the terms of the image sum, whose contributions either have the same 
 sign 
and add up, resulting in local maxima and minima for negative gaps, or have different signs and cancel, resulting in a function without oscillations for positive gaps. In both cases, however, the overall trend of the transition rate is to increase as more glitches $\Delta\tau_n/\ell$ are reached, or the singularity at $r=0$ is approached by the detector.

The transition rate for negative and large enough energy gaps becomes more intricate as $\Delta\tau/\ell$ increases due to the fact that the contribution of the $n$-th term of the image sum, which becomes significant for detection times larger than $\Delta\tau_n/\ell$, has a quasiperiod different from that of previous terms. 
Another remarkable trait for negative gaps is the alternating order of the local maxima and minima in the transition rate after each $\Delta\tau_n/\ell$ as  $|E\ell|$ increases. 
Finally, another feature that can be retrieved from Fig.~\ref{fig:varying_gap_z_1} is that for smaller $|E\ell|$, the earlier the onset of the deviation of the transition rate (solid curves) from the zeroth term (dotted curves). Such deviation occurs at detection times for which the detector has not yet reached the event horizon in all the cases shown.

We see in Fig.~\ref{fig:varying_gap_z_m1}, and in the previous subsection, that the discrepancy between the transition rate and the pure AdS$_3$ term is more noticeable for the boundary condition $\zeta=-1$ and is present for most of the detection time range.

\subsection{Transition rate for different masses}\label{sec:diff_masses}

\begin{figure*}[t]
    \centering
    \includegraphics[width=0.49\linewidth]{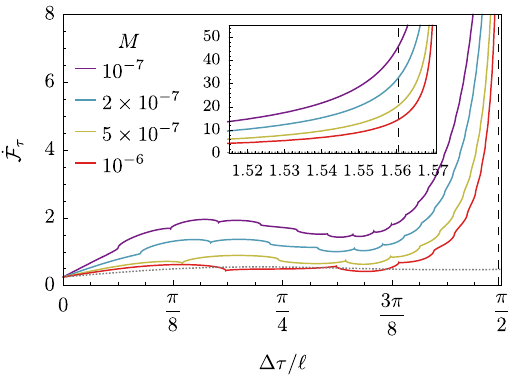}
    \includegraphics[width=0.49\linewidth]{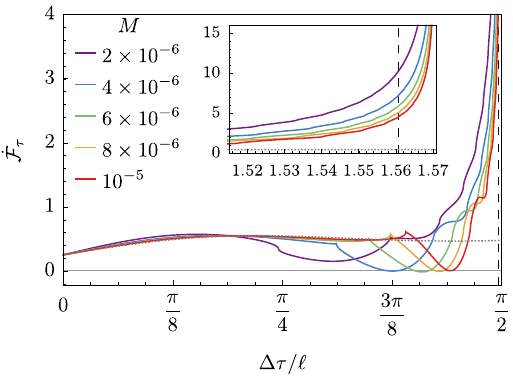}
    \includegraphics[width=0.49\linewidth]{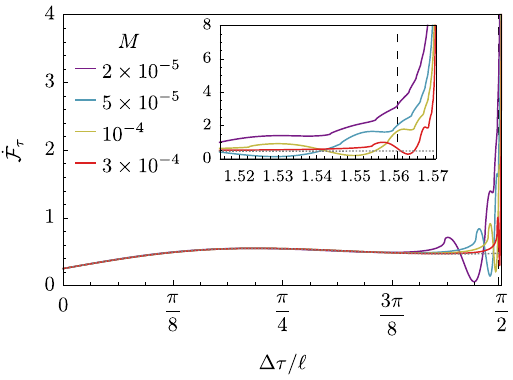}
    \includegraphics[width=0.49\linewidth]{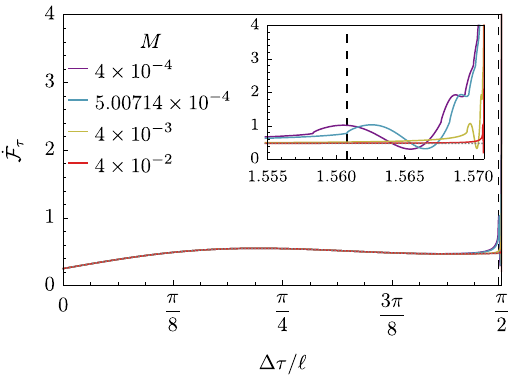}
    \caption{Transition rate $\dot{\mathcal{F}}_\tau$ as a function of $\Delta\tau/\ell$ for energy gap $E\ell=-5$, boundary condition $\zeta=1$ and black hole mass $M\le10^{-6}$ (top-left), $2\times10^{-6}\le M \le 10^{-5}$ (top-right), $2\times10^{-5}\le M \le 3\times10^{-4}$ (bottom-left) or $ M \ge 4\times10^{-4}$ (bottom-right). The corresponding zeroth term (the AdS$_3$ case) 
    is also shown with grey, dotted curves. The horizon crossing is shown with a dashed vertical line at $\Delta\tau/\ell=\arccos{(0.01)}$. The total detection time ranges from 0 to $\pi/2$ in the main plots and from $1.51$ to $\pi/2$ in the insets, except in the last inset, where it ranges from $1.555$ to $\pi/2$. $n=11236$, 7946, 5026, 3553, 2513, 1779, 1454, 1257, 1124, 796, 506, 356, 252, 218, 195, 70 and 25 terms of the image sum are used in order of increasing mass.}
    \label{fig:different_masses}
\end{figure*}

With fixed energy gap $E\ell=-5$, $q=100$, $\tau_0=0$ and $\zeta=1$, we calculate the transition rate for a range of black hole mass,
plotting the results in Fig.~\ref{fig:different_masses}. We
 find that the rate is quite sensitive to  the magnitude of the black hole mass.  
More specifically, the transition rate calculated with $M=10^{-7}$, the smallest mass we explored, is punctuated by sharp local maxima and minima, corresponding to several glitches $\Delta\tau_n/\ell$ occurring before the horizon crossing. For this particular mass, the image sum was truncated at the $n=11236$ term. We observe that, as the black hole mass increases, the first glitch $\Delta\tau_1/\ell$ appears at a longer detection time, the number of glitches $\Delta\tau_n/\ell$ before the horizon crossing decreases, and fewer terms in the image sum are necessary to calculate a converged transition rate. We also observe that there is a clear difference between the transition rate $\dot{\mathcal{F}}_\tau$ and the zeroth term $\dot{\mathcal{F}}_\tau^{n=0}$ for $M=10^{-7}$. However, this deviation from the zeroth term for $\Delta\tau/\ell\lesssim\Delta\tau_{1}/\ell$ becomes negligible as the black hole mass gets bigger.  These results suggest that the smaller the mass of the black hole, the sooner a BTZ black hole spacetime can be distinguished from AdS$_3$ spacetime with the same constant curvature.

In Fig.~\ref{fig:different_masses} (top-left), the broader picture is similar across the range of mass $10^{-7} \le M \le 10^{-6}$. The transition rate starts increasing, as observed for $E/\ell<0$ in the previous subsection, and reaches a peak at $\sim\pi/8$. 
It then attains a local minimum at $\sim5\pi/16$ and 
later grows rapidly, increasing significantly as the detector crosses the horizon. 
For masses $M\ge2\times10^{-6}$, the transition rate has a close resemblance to the $n=0$ term until reaching the first glitch $\Delta\tau_1/\ell$. 

For the range of mass $2\times10^{-6} \le M \le 10^{-5}$ (Fig.~\ref{fig:different_masses}, top-right), the transition rate has a pronounced dip after the first glitch. By contrast, it has a prominent peak after the first glitch for $M\ge2\times10^{-5}$, Fig.~\ref{fig:different_masses} (bottom-left and bottom-right). 
The dip (or peak) is enhanced as the mass increases. 

In Fig.~\ref{fig:different_masses} (bottom-right), we find that the first glitch in the transition rate is at the horizon crossing for $M\approx5.00714\times10^{-4}$. For larger masses, the peak after the first glitch is entirely inside the horizon, and the transition rate has the same shape as for small masses, but the oscillations are squeezed closer to the singularity.
However, we still observe a noticeable difference in the transition with respect to the $n=0$ term, even for masses as big as $M=4\times10^{-3}$. It is remarkable that this difference is noticeable for detection times for which the detector has not yet reached the event horizon.

For other boundary conditions ($\zeta=0,-1$; see Fig.~\ref{fig:different_masses_zm1}), the peak at $\pi/8$ and the dip at $5\pi/16$ get a larger amplitude, becoming more apparent. Moreover, the dip at $5\pi/16$ reaches negative values for small masses and/or becomes a global minimum. On the other hand, the transition rate with these boundary conditions gets a positive shift that increases with the detection time for $\Delta\tau\gtrsim3\pi/8$. As a result, the dips become subtler, and the peaks are enhanced for a long enough detection time with these boundary conditions.

\subsection{Transition rate varying  $q$}\label{sec:diff_qs}

\begin{figure}[t]
    \centering
    \includegraphics[width=\linewidth]{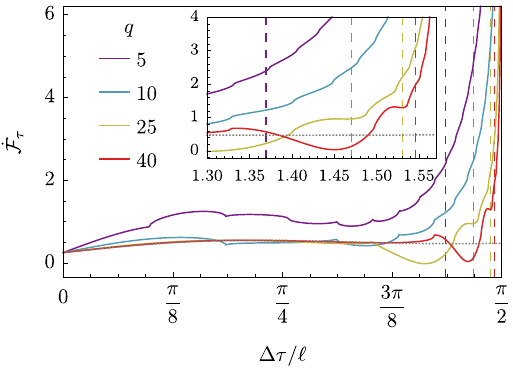}
    \includegraphics[width=\linewidth]{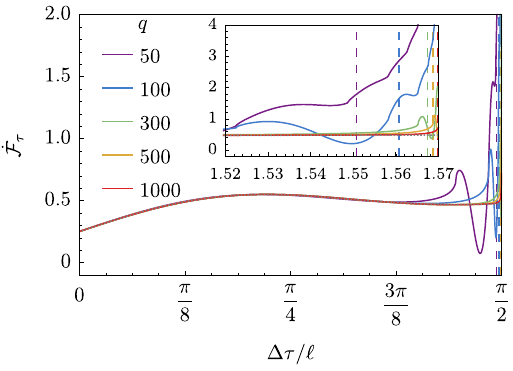}
    \caption{Transition rate $\dot{\mathcal{F}}_\tau$ as a function of $\Delta\tau/\ell$ for $\tau_0=0$, $E\ell=-5$, $\zeta=1$, $M=10^{-4}$ and varying $q$. The image sum in the transition was truncated at the term $n=451$, 429, 400, 385, 378, 356, 321, 305 and 281  in order of increasing $q$. The different horizon crossings, at $\Delta\tau/\ell=\arccos(1/q)$, are indicated with vertical dashed lines in the same colour as the transition rates for each value of $q$.}
    \label{fig:differentq}
\end{figure}

In Fig.~\ref{fig:differentq}, we explore the transition rate with fixed $E\ell=-5$, $\tau_0=0$, $\zeta=1$, $M=10^{-4}$ and varying $q$, the parameter that determines the detector's initial radial position with respect to the black hole's event horizon. We observe that fixing the black hole mass $M$ and varying $q$ has the same effect on the transition rate as fixing $q$ and varying $M$ (see Fig.~\ref{fig:different_masses}). 
Namely, the first glitch is located at a longer detection time, the number of glitches before the horizon crossing decreases, fewer terms are needed to calculate the image sum, and the deviation from the pure AdS$_3$ term is reduced as the parameter $q$ increases.
This similar effect is expected since an increasing function of $q$ is multiplying an increasing function of $M$ in the second term of $K_n$ of Eq.~\eqref{eq:K_n}, and the parameters $q$ and $M$ affect the transition rate only through $K_n$. 
However, increasing $q$ also increases the total proper time for the detector to reach the event horizon. 
Finally, the results in Fig.~\ref{fig:differentq} imply that the closer the detector is initially located with respect to the event horizon, the sooner it can discern between being in an AdS$_3$ spacetime and a BTZ black hole.

\begin{figure*}[t]
    \centering
    \includegraphics[width=0.7\textwidth]{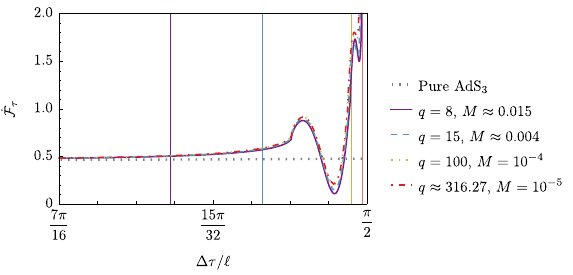}
    \caption{Transition rates $\dot{\mathcal{F}}_\tau$ as a function of $\Delta\tau/\ell$ for $\tau_0=0$, $E\ell=-5$ and $\zeta=1$, varying both $M$ and $q$ such that the first glitch occurs at the same time. The horizon crossing, $\Delta\tau/\ell=\arccos(1/q)$, is indicated with a vertical solid line in the same colour as the transition rate for each value of $q$.}
    \label{fig:same_rate_diff_q_and_M}
\end{figure*}

Given that a short initial distance paired with a large mass produces a transition rate that resembles that of a large initial distance paired with some small mass, we investigate combinations of $q$ and $M$ that produce similar transition rates. For this purpose, we compare in Fig.~\ref{fig:same_rate_diff_q_and_M} transition rates that have the first glitch, $\Delta\tau_1/\ell$, at the same total detection time. One can find $M_2$ and $q_2$ that will produce the same first glitch in the transition rate as $M_1$ and $q_1$; however, the other glitches, $\Delta\tau_n/\ell$ for $n>1$, are not guaranteed to be equal. This parameter calibration results in transition rates that are qualitatively the same, regarding the positions of their peaks and dips, at least for a detection time close to $\Delta\tau_1/\ell$. As a consequence, the peaks and dips (or glitches) can appear in the transition rate before or after (and close or far from) the proper time at which the detector reaches the event horizon, as the horizon crossing depends on $q$ through  $\Delta\tau/\ell=\arccos{(1/q)}$. 
This observation implies that, without independent determination of the black hole mass, the approximate positions of the peaks or dips in the transition rate cannot be used to establish an early warning system   indicating when Alice, an observer freely falling with a detector into a BTZ black hole, will cross the event horizon. However, she could still discern a BTZ black hole spacetime from an AdS$_3$ spacetime based on the presence or absence of these peaks, albeit possibly after the horizon is crossed.

In the case that Alice can measure the transition rate with infinite precision,
then not only   could she  distinguish the difference between AdS$_3$ spacetime and a BTZ black hole spacetime, but she could also determine, at any time, when she is crossing the event horizon by estimating the value of $q$ and the black hole's mass $M$, distinguishing between any of the cases presented in Fig.~\ref{fig:same_rate_diff_q_and_M}. Therefore, Alice could establish an early warning system that can alert her that she is entering the black hole by ideally measuring the transition rate of the detector.

 We now consider a scenario where these ideal measurements of the detector's transition rate are inaccessible, and only the time at which each glitch occurs ($\Delta\tau_n/\ell$) is available. In this situation, Alice can still estimate the mass of the black hole and the location of its event horizon. Both $M$ and $q$ can be determined by having information about two of the glitches in the transition rate, as described by equations \eqref{eq:K_n} and \eqref{eq:glitches}. Let us assume that Alice can measure the first and second glitches. With this information, she can establish a ``weaker'' early warning system given that the second glitch happens before the detector reaches the event horizon, that is,
\begin{equation}
    \Delta\tau_2 / \ell < \arccos{(1/q)}.
    \label{eq:condition_early_warning}
\end{equation}
For fixed $q$, \eqref{eq:condition_early_warning} is satisfied when 
\begin{equation}
    0<\sqrt{M}<\frac{1}{2\pi}\operatorname{arcsinh}{\left(\frac{q-1}{2q^2} \right)^{1/2}},
\end{equation}
and for fixed mass $M$, \eqref{eq:condition_early_warning} is satisfied if
\begin{equation}
    0<\sqrt{M}<\frac{1}{2\pi}\operatorname{arcsinh}{\frac{1}{2\sqrt{2}}} \quad \text{and} \quad q_-<q<q_+,
\end{equation}
where
\begin{equation}
    q_{\pm} = \frac{1\pm \sqrt{1-8\sinh^2(2\pi\sqrt{M})}}{4\sinh^2(2\pi\sqrt{M})}.
\end{equation}
This means that the weaker early warning system is constrained to work for a black hole mass below the critical value 
 $M_* = 
\left[1/(2\pi)\operatorname{arcsinh}{\left(1/(2\sqrt{2})\right)}\right]^2
\approx3.0425\times10^{-3}$, and 
within a limited range of the detector's initial position.

\section{Conclusion}
\label{sec:conclusions}

We have numerically calculated the transition rate of an Unruh-DeWitt detector 
coupled to a  conformal scalar field as the detector falls radially
into a spinless BTZ black hole.
We have been able to explore a broad range of values of the black hole mass,  the detector's energy gap and initial position,
and scalar field boundary conditions at asymptotic infinity. The
problem is technically tractable 
due to the relatively simple calculation of the Wightman function of the BTZ black hole as the image sum of the corresponding AdS$_3$ spacetime Wightman function.
We  found that the transition rates have the same qualitative behaviour across  different  boundary conditions (though differing quantitatively), 
and we have observed that the transition rate differs significantly for detectors with positive and negative energy gaps.

We discovered that  the detector's transition rate is nondifferentiable at certain discrete values of the detector's proper time; we refer to these points of nondifferentiability   as glitches.
There is one glitch in the transition rate for each term ($n\ge1$) of the image sum in the BTZ Wightman function, and it corresponds to the  proper time of the detector at  which the integrand in the transition rate's final expression diverges. As a result, the transition rate is not smooth at these glitches.
The contribution from any term ($n\ge1$) to the transition rate is negligible for detection times shorter than the $n$-th glitch but rapidly becomes prominent for detection times larger than the glitch, exhibiting oscillations that are determined by the magnitude and the sign of the detector's energy gap. Consequently, the transition rate for a detector freely infalling  into a BTZ black hole resembles that of a detector in an AdS$_3$ spacetime for detection times shorter than the first ($n=1$) glitch. After the first glitch, the transition rate presents a richer and more complex behaviour.

The glitches are fully determined by the black hole's mass and the detector's initial position relative to the event horizon. Adjusting these parameters changes the occurrence of glitches in the transition rate. We   found that increasing the black hole's mass and the detector's initial position results in the BTZ transition rate resembling the AdS$_3$ transition rate for a longer period of time (as the first glitch appears at a later detection time) and in the number of glitches before the horizon being smaller (requiring fewer terms to obtain a converging transition rate).
Conversely, for smaller  mass or closer initial position, the transition rate starts to differ from that of pure AdS$_3$ at shorter detection times.

Our study was motivated by the observation that the response function of a detector interacting with the Hartle-Hawking(-Israel) state and freely falling in a Schwarzschild black hole has a non-monotonic behaviour and presents a small but discernible local extremum as the detector's crosses the event horizon~\cite{Ng2022}.
Our results indicate that
a similar phenomenon is present for
the BTZ black hole, but with much richer behaviour than in the Schwarzschild   case. Specifically, more than one local extrema can be present across the horizon for certain ranges of the black hole's mass, the detector's initial position and energy gap. The detection time at which these extrema appear in the transition rate will be determined by the black hole's mass and the detector's relative position with respect to the event horizon. Meanwhile, the number and amplitude of these extrema are determined by the magnitude and sign of the detector's energy gap.

Finally, we have suggested that an observer with a detector that is freely falling can discern between being in an AdS$_3$ spacetime and a BTZ black hole spacetime by measuring the detector's transition rate. Moreover the observer would be able to know, in their proper time, when they are crossing the black hole's event horizon, having in this way a sort of an early warning system.
We also proposed a second version of the warning system that would depend only on the time at which the first two glitches happen. If the second glitch occurs before the time of the horizon crossing, then an observer can know the position of the event horizon before they cross it. This weaker version of the early warning system is  valid only for masses smaller than a certain critical mass and for a limited range of the detector's initial position.

Our results are somewhat limited insofar as they do not allow for a direct comparison between the Schwarzschild and BTZ black holes. One limitation is that the detector's initial conditions are  different.
Another is that the BTZ spacetime has constant curvature whereas the Schwarzschild case does not. 
Yet another limitation is that  the response function was calculated for  the Schwarzschild spacetime \cite{Ng2022}, whereas we computed the transition rate for the  BTZ spacetime. We leave the task of comparing both response functions instead of the transition rate subject to future work.

Our results suggest a number of new research directions. One is to consider the effects of hidden topology --
 for example, 
 comparing the transition rate of an infalling detector for a BTZ black hole to its  corresponding geon counterpart.
Adding effects of rotation should be straightforward, since the curvature remains constant.  
Another interesting problem is to study what happens to two maximally entangled detectors when one (or both) falls into a black hole. 
More challenging problems will involve  black hole spacetimes that do not have constant curvature.

\emph{Note added during proof.} After the completion of this work, we became aware of \cite{Conroy:2021aow}, which considers the transition rate of selected families of trajectories in selected quantum states in four-dimensional Bertotti-Robinson spacetime. We thank Peter Taylor for bringing this work to our attention.

\section*{Acknowledgments}
We thank Suprit Singh for bringing the work in 
\cite{Smerlak:2013sga} to our attention
and an anonymous referee for helpful comments. 
This work was supported in part by the Natural Sciences and Engineering Research Council of Canada.
MRPR acknowledges that this research was undertaken thanks in part to the funding from the Mike and Ophelia Lazaridis Graduate Fellowship.
The work of JL was supported by United Kingdom Research and Innovation Science and Technology Facilities Council [grant number ST/S002227/1]. For the purpose of open access, the authors have applied a CC BY public copyright licence to any Author Accepted Manuscript version arising. 

\appendix*

\section{Additional results}\label{sec:appendix}

\begin{figure}[t]
    \centering
    \includegraphics[width=\linewidth]{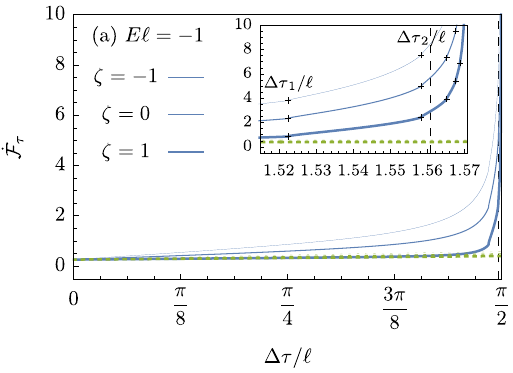}
    \includegraphics[width=\linewidth]{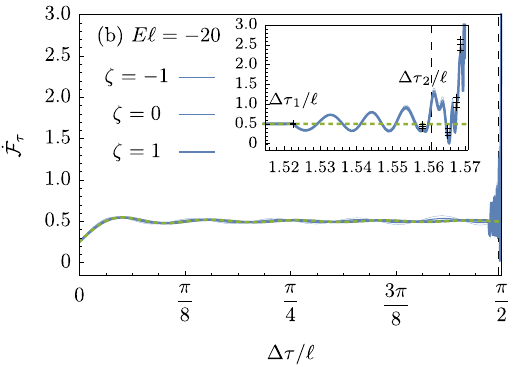}
    \caption{Zeroth term $\dot{\mathcal{F}}_\tau^{n=0}$ (green, dotted curves) and transitions rate $\dot{\mathcal{F}}_\tau$ (blue, solid curves) for the same parameters as Fig.~\ref{fig:comparingtojorma} and (a) $E\ell=-1$ or (b) $E\ell=-20$ calculated with $n=388$ or $n=320$ terms of the image sum, respectively.
    \label{fig:boundary_conditions_negative_gap}
    }
\end{figure}
\begin{figure}[t]
    \centering
    \includegraphics[width=\linewidth]{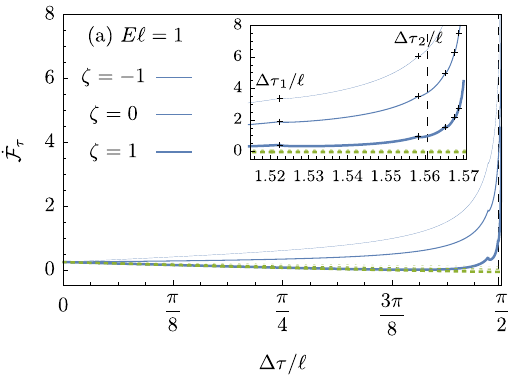}
    \includegraphics[width=\linewidth]{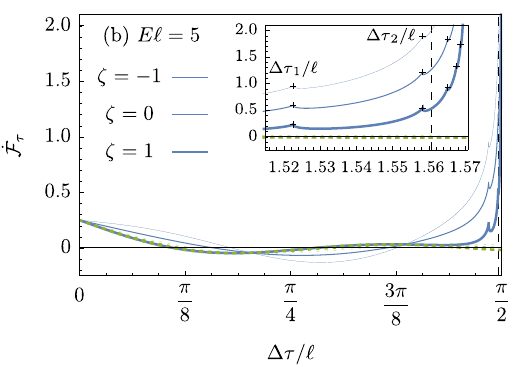}
    \caption{
        Zeroth term $\dot{\mathcal{F}}_\tau^{n=0}$ (green, dotted curves) and transitions rate $\dot{\mathcal{F}}_\tau$ (blue, solid curves) for the same parameters as in Fig.~\ref{fig:comparingtojorma} and (a) $E\ell=1$ or (b) $E\ell=5$ calculated with $n=388$ or $n=356$ terms of the image sum, respectively.
    }
    \label{fig:boundary_conditions_positive_gap}
\end{figure}

In this appendix, we present additional results that complement the discussion in Section \ref{sec:results}. 
In particular, we show the transition rate as a function of the detector's proper time $\Delta\tau/\ell$ for the three boundary conditions of the Wightman function, $\zeta=-1,\,0,$ and $1$, with the same parameters used in Fig.~\ref{fig:comparingtojorma} but different energy gaps, specifically $E\ell=-1$ and $E\ell=-20$ in Fig.~\ref{fig:boundary_conditions_negative_gap}, and $E\ell=1$ and $E\ell=5$ in Fig.~\ref{fig:boundary_conditions_positive_gap}. 

We first discuss the transition rate for detection times shorter than the first glitch $\Delta\tau_1/\ell$ (with $\Delta\tau_1/\ell\approx 1.52257$ for $q=100$ and $M=10^{-4}$).
We observe in Figs. \ref{fig:boundary_conditions_negative_gap} and \ref{fig:boundary_conditions_positive_gap} that the Dirichlet boundary condition ($\zeta=1$) bears a strong resemblance to the corresponding pure AdS$_3$ transition rate, that is, the $n=0$ term,  across the different energy gaps shown. This is due to partial cancellations in the terms of equation \eqref{eq:transition_rate}.
For the other boundary conditions, $\zeta=0$ and $\zeta=-1$, it can be argued the transition rate presents the same qualitative behaviour as the pure AdS$_3$ transition rate. However, there is a quantitative dissimilarity for these boundary conditions that increases as the magnitude of the energy gap $|E\ell|$ decreases. We see in Fig.~\ref{fig:comparingtojorma}(a) and Fig.~\ref{fig:boundary_conditions_positive_gap}(b) that the difference between these transitions rate becomes noticeable for $|E\ell|=5$.

We now discuss the transition rate near and beyond the first glitch, which is shown in more detail in the insets of Figs.~\ref{fig:boundary_conditions_negative_gap} and \ref{fig:boundary_conditions_positive_gap}. As pointed out before, the transition rate qualitatively differs from the pure AdS$_3$ case (the zeroth term) for detection times near and beyond the first glitch, $\Delta\tau/\ell \gtrsim \Delta\tau_1/\ell$, across the different boundary conditions and energy gaps. This difference is explained in the discussion of Fig.~\ref{fig:individual_contributions}. While the transition rate calculated with a negative gap oscillates with a pseudo-period proportional to the magnitude of the gap, as observed in Fig.~\ref{fig:boundary_conditions_negative_gap}, the transition rate calculated with a positive gap has dips after each glitch but does not oscillate, as observed in Fig.~\ref{fig:boundary_conditions_positive_gap}.

Another noteworthy result is that the transition rates for a freely falling detector in a BTZ black hole calculated with a negative energy gap, as in Fig.~\ref{fig:boundary_conditions_negative_gap}, also exhibit the features of the transition rate in Fig.~\ref{fig:comparingtojorma}(b), that is, they present oscillations that resemble the local extremum obtained in the Schwarzschild black hole case \cite{Ng2022}. 

Since the discussion of the results in the main text was limited to the $\zeta=1$ case  when studying the transition rate for different detector energy gaps in Section~\ref{sec:diff_gaps}, black hole mass in Section~\ref{sec:diff_masses}, and initial parameter $q$ in Section~\ref{sec:diff_qs}, we   present here the transition rate calculated with the boundary condition $\zeta=-1$ for completeness. The transition rate for $\zeta=0$ (not shown) is always between these two results. 

\begin{figure*}[ht]
    \centering
    \includegraphics[width=0.49\textwidth]{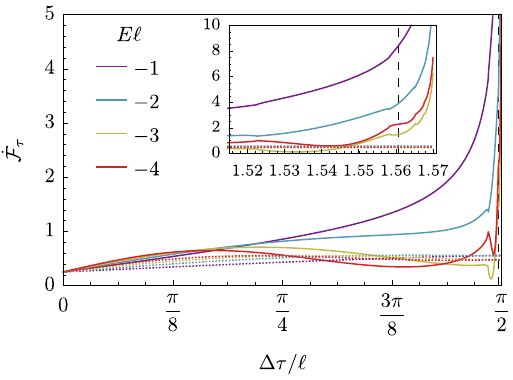}
    \includegraphics[width=0.49\textwidth]{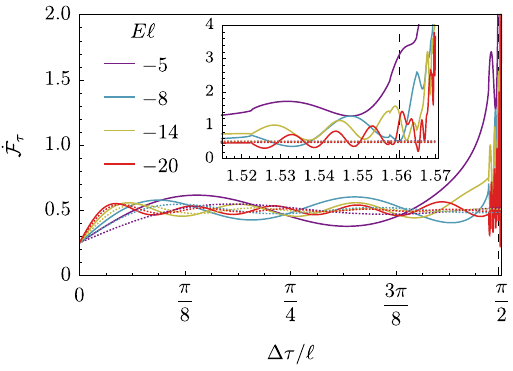}
    \includegraphics[width=0.49\textwidth]{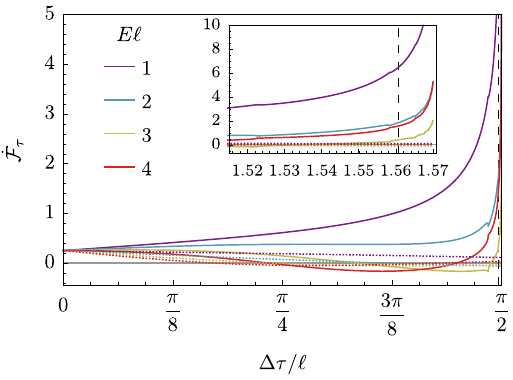}
    \includegraphics[width=0.49\textwidth]{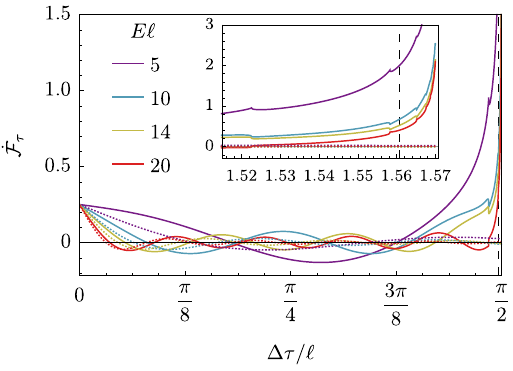}
    \caption{
    Transition rate $\dot{\mathcal{F}}_\tau$ as a function of $\Delta\tau/\ell$ for the same parameters as Fig.~\ref{fig:varying_gap_z_1} but boundary condition $\zeta=-1$. The pure AdS$_3$ (zeroth term $\dot{\mathcal{F}}^{n=0}_\tau$) for each gap is shown with dotted curves of the same colour.
    }
    \label{fig:varying_gap_z_m1}
\end{figure*}

In Fig.~\ref{fig:varying_gap_z_m1}, we show the transition rate varying the detector's energy gap and using the same parameters as in Fig.~\ref{fig:varying_gap_z_1} but with boundary condition $\zeta=-1$. While the overlap between the zeroth term (dotted curves) and the transition rate (solid curves) lasts up to detection times close to the first glitch in the results for the boundary condition $\zeta=1$ in Fig.~\ref{fig:varying_gap_z_1}, we observe that there is little to no overlap between the zeroth term and the transition rate for $\zeta=-1$ across the different energy gaps in Fig.~\ref{fig:varying_gap_z_m1}. This observation agrees with what was discussed in the sections regarding the transition rate for the different boundary conditions.
Overall, the changes produced in the transition rate with $\zeta=-1$, as the magnitude and sign of the energy gap vary, are similar to the ones discussed for the transition rate with $\zeta=1$.

\begin{figure*}[ht]
    \centering
    \includegraphics[width=0.49\linewidth]{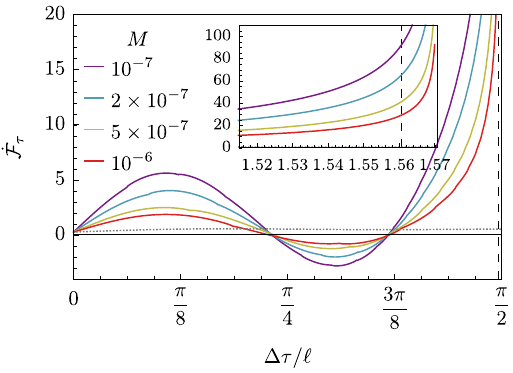}
    \includegraphics[width=0.49\linewidth]{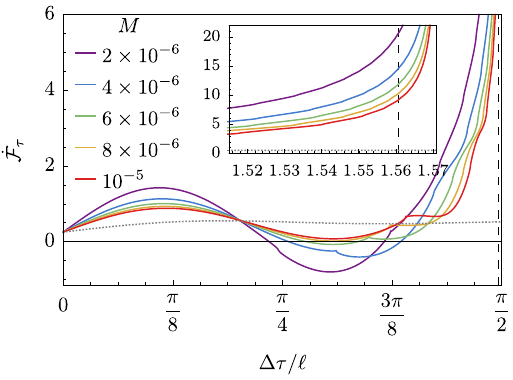}
    \includegraphics[width=0.49\linewidth]{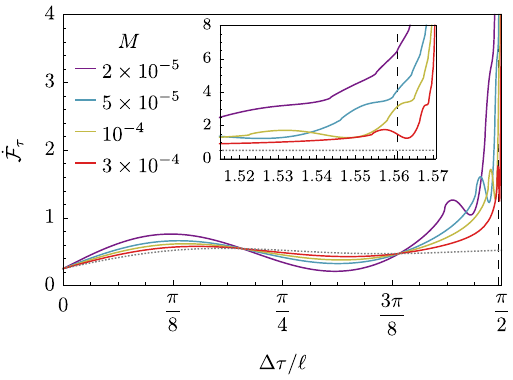}
    \includegraphics[width=0.49\linewidth]{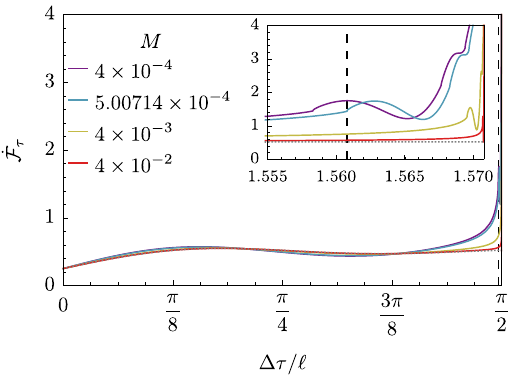}
    \caption{
    Transition rate $\dot{\mathcal{F}}_\tau$ as a function of $\Delta\tau/\ell$ for the same parameters as Fig.~\ref{fig:different_masses} but boundary condition $\zeta=-1$.
    The corresponding zeroth term (the AdS$_3$ case) is also shown with grey, dotted curves.
    }
    \label{fig:different_masses_zm1}
\end{figure*}

In Fig.~\ref{fig:different_masses_zm1}, we depict the transition rate for a range of black hole mass. We use the same fixed parameters as in Fig.~\ref{fig:different_masses} but boundary condition $\zeta=-1$.
We can observe that the magnitude of the oscillations with local maximum at $\pi/8$ and minimum at $5\pi/16$ is considerably greater for the $\zeta=-1$ case. As a result, the sharp peaks corresponding to the glitches in the transition rate are less noticeable in Fig.~\ref{fig:different_masses_zm1} than in Fig.~\ref{fig:different_masses}. As discussed in the main text, for $\Delta\tau/\ell \gtrsim 3\pi/8$, a positive shift for $\zeta=-1$ in the transition rate that increases with the detection time causes the peaks to become more apparent and the dip to become more subtler when compared to the $\zeta=1$ case.

\begin{figure}[ht]
    \centering
    \includegraphics[width=\linewidth]{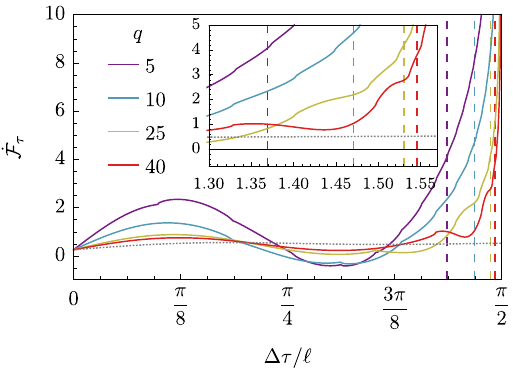}
    \includegraphics[width=\linewidth]{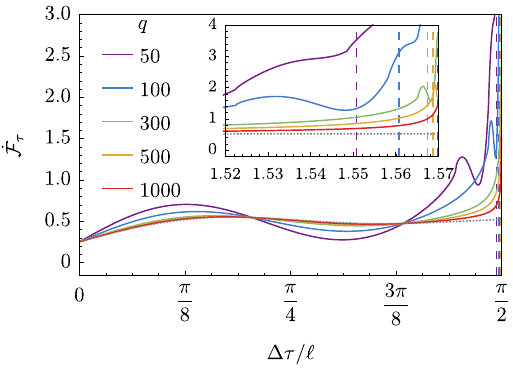}
    \caption{Transition rate $\dot{\mathcal{F}}_\tau$ as a function of $\Delta\tau/\ell$ for the same parameters as Fig.~\ref{fig:differentq} but boundary condition $\zeta=-1$.}
    \label{fig:differentq_zm1}
\end{figure}

 Finally, in Fig.~\ref{fig:differentq_zm1}, we present the transition rate by varying $q$, calculated with the same parameters as in Fig.~\ref{fig:differentq} but for $\zeta=-1$. As also observed in the transition rate calculated  for variable black hole mass, the effect of the changing the boundary condition from $\zeta=1$ to $\zeta=-1$ is that the amplitude of the first peak and dip located at $\Delta\tau/\ell=\pi/8$ and $\Delta\tau/\ell=5\pi/16$ gets larger.

\newpage 
$\phantom{xxx}$ 
\newpage

\bibliography{ref}

\end{document}